\documentclass[
showpacs,
nofootinbib,
amsmath,amssymb,
prd,
twocolumn
]{revtex4-1}

\usepackage{graphicx}
\usepackage[all,arc,knot,matrix,color]{xy}
\usepackage{braket}
\usepackage{natbib}
\usepackage{hyperref}
\usepackage{CJK}

\begin{document}

\begin{CJK*}{UTF8}{bkai}
\title{Contextual extensions of quantum gravity models}

\author{Xiao-Kan Guo~(郭肖侃)}
 \email{kankuohsiao@whu.edu.cn}
\affiliation{Department of Physics, Beijing Normal University, Beijing 100875, China}
\date{\today}

\begin{abstract} 
We present a simple way of incorporating  the  structure of contextual extensions into quantum gravity models. The contextual extensions of $C^*$-algebras, originally proposed for contextual hidden variables, are  generalized to the cones indexed by the contexts and their limit  in a category. By abstracting the quantum gravity models as functors, we study the contextual extensions as the categorical limits of these functors in several quantum gravity models. Such contextual extensions of quantum gravity models are useful for building topos-theoretic models of quantum gravity.
\end{abstract}



\maketitle
\end{CJK*}

\section{Introduction}
In recent years, there is a surge of interest in exploring  the entanglement structures  in quantum gravity. In contrast to the traditional studies in quantum gravity that focus on the quantization and the observational effects, considering quantum entanglement reveals the truly quantum features a quantum theory of  gravity ought to possess.

In standard quantum theory, the quantum entanglement  between two space-like separated systems allows nonlocal correlations, as  highlighted by the Bell nonlocality \cite{Bell64}. The spatial nonlocality of quantum entanglement can be understood in a contextually local way, since quantum theory is contextual as is shown by the Kochen-Specker theorem \cite{KS67}. However, neither the Bell nonlocality nor the contextuality plays a major role in most models of quantum gravity. On the one hand, a  quantized spacetime is  expected to be drastically different from the classical spacetime, so that the space-like Bell nonlocality in no longer meaningful in  quantum gravity.  On the other hand, there is no need in quantum gravity for quantum  measurements by an external observer, as is well-known in the cosmological treatment. Therefore there is no simple way in quantum gravity to talk about the contexts for co-measurable observables. Nevertheless, these properties should have their roles in quantum gravity, as long as the quantum theory of gravity is still a standard quantum theory. 

In various quantum gravity  models , one usually uses the entanglement entropy to probe the cohesiveness of (either classical or quantum) spacetime composed of different regions  \cite{BM14}, but the entropy along cannot distinguish  a truly quantum theory from a theory of local hidden variables.  A study of Bell inequality might help us to test whether a quantum gravity model is truly quantum. (See e.g. \cite{CSZ19}  for a holographic model of Bell inequality). Meanwhile, in many theoretical studies of quantum gravity one considers the operators for  quantum observables and their spectra, but does not restrict these operators to be compatible to  have definite eigenvalues. However, we know from the Kochen-Specker theorem that the simultaneous global valuation is impossible if the Hilbert-space dimension is greater than 2.
The contextual structures could emerge   when we compute the probabilities for events or the transition amplitudes for processes in quantum gravity. (See e.g. \cite{Ish97} for the contextual structure in the consistent history formulation of quantum cosmology.)

In this short paper, we study the  structures of contextual extensions in some  models of quantum gravity. The  motivation for considering contextual structures comes from the works about finding the suitable ``quantum'' logical structure of quantum gravity. In the consistent-history quantum cosmology \cite{Ish97}, as well as in the quantum causal sets \cite{Mar00,Rap02}, the logical structure is argued to be described by topos theory. The subsequent developments result in the topos quantum theory \cite{Flo13,Flo18} which naturally incorporates the contextual structure, but
  the relations between the contextual structures and   quantum gravity or even spacetime physics  become more obscure. In  other attempts starting with lattice categories  in quantum gravity models, for example \cite{Cra07,Guo18}, one is finally  led to the topos-theoretic structures. We therefore hope to include the contextual structures into quantum gravity models in  such a way that they can be related to the topos structure. To this end, we revisit the contextual extensions of $C^*$-algebraic models of quantum theory, as is developed in \cite{Dur91,Dur92} for contextual hidden variable theories. This kind of  contextual extensions are the extensions of theories with {\it existing} contexts, which is acceptable since  a quantum theory of gravity is expected to be within the standard quantum formalism.
  
By restating the contextual extensions of $C^*$-algebras in terms of category theory, we are able to formulate the contextual extensions in various, both nonperturbative and perturbative, models of quantum gravity. As examples, we consider the contextual extensions in the algebraic formulation of group field theory, in the lattice theory of spin foam models and in the  Batalin-Vilkovisky (BV) formulation of perturbative quantum gravity. These examples imply a general (higher) topos structure of the contextually extended quantum gravity models.
  
We start   in \S\ref{II} by recalling the contextual extensions of algebraic quantum theories. We also show that the contextual extensions naturally exist in topos quantum theory. In \S\ref{III}, we consider the contextual extensions of quantum gravity models. \S\ref{IV}  concludes.
\section{Contextual extensions }\label{II}
In this section we first review the contextual extensions of $C^*$-algebras both for quantum mechanics and quantum field theories \cite{Dur91,Dur92}. Then we show that the contextual extensions  can be naturally found in  topos quantum theory \cite{Flo13,Flo18}.
\subsection{Contextual extensions of algebraic quantum theories}
In  algebraic  quantum theory, a quantum state $\varphi$ is a positive linear functional on the unital $C^*$-algebra $\mathcal{A}$ of quantum observables. The Gel'fand-Naimark-Segal (GNS) construction on $\mathcal{A}$ induced by $\varphi$ gives the GNS representation $(\mathcal{H},\pi,\ket{\Omega})$ with $\pi:\mathcal{A}\rightarrow\mathcal{B}(\mathcal{H})$ being a *-homomorphism to the bounded operators on the Hilbert space $\mathcal{H}$ and $\ket{\Omega}\in\mathcal{H}$ a cyclic separating vector. For $A\in\mathcal{A}$, we have $\varphi(A)=\braket{\Omega|\pi(A)|\Omega}$.

The measurement contexts of $\mathcal{A}$ are defined as a family ${\bf V}$ of commutative $C^*$-subalgebras of $\mathcal{A}$. 
 Let us denote by $i_{\mathcal{V}}:\mathcal{V}\rightarrow \mathcal{A}$ the inclusion of subalgebras.
 The {\it contextual extension} of $\mathcal{A}$, in a general sense, consists of a finer $C^*$-algebra $\mathcal{A}'$, a surjective $*$-homomorphism $\phi:\mathcal{A}'\rightarrow\mathcal{A}$, and a family of $*$-homomorphisms $\{\iota_\mathcal{V}:\mathcal{V}\rightarrow\mathcal{A}'\}_{\mathcal{V}\in{\bf V}}$ as subalgebra inclusions  such that $\phi\iota_\mathcal{V}=i_\mathcal{V}$. 
 Diagrammatically, we have
\begin{equation}\label{1}
\vcenter{\xygraph{
!{<0cm,0cm>;<1cm,0cm>:<0cm,1cm>::}
!{(0,0) }*+{\mathcal{A}}="a"
!{(1,1) }*+{\mathcal{V}}="b"
!{(2,0) }*+{\mathcal{A}'}="c"
"b":"a"_{i_\mathcal{V}}"b":"c"^{\iota_\mathcal{V}}"c":"a"^{\phi}
}}\end{equation}
In category-theoretic terms, we consider the category $\mathsf{V}$ formed by those $\mathcal{V}\in{\bf V}$ and as morphisms  the injective $*$-homomorphisms,  and the  category $\mathsf{A}_*$ of  $C^*$-algebras (including those $\mathcal{V}$s) with the morphisms being the $*$-homomorphisms in the reverse direction ($\phi^{-1}$). 
Then
we have the cone \eqref{1} over the a diagram defined by the functor $I:\mathsf{V}\rightarrow\mathsf{A}_*$.

Taking the limiting cone $\lambda$ of \eqref{1} we have the limit object $\mathcal{A}_{\text{ctx}}\in\text{Obj}(\mathsf{A}_*)$, and $\i_\mathcal{V}:\mathcal{V}\rightarrow\mathcal{A}_{\text{ctx}}$ such that $\lambda\circ\i_\mathcal{V}=\iota_\mathcal{V}$ for any $\mathcal{V}$. Due to the universality of limit,  all possible measurement contexts in ${\bf V}$ can be included into $\mathcal{A}_{\text{ctx}}$; in other words, $\mathcal{A}_{\text{ctx}}$ is the ``biggest'' contextual extension of  $\mathcal{A}$.
For any $A\in\mathcal{A}$ in a context $\mathcal{V}$, we collect them into a new element $(A,\mathcal{V})$; in a fixed context $\mathcal{V}$, the $*$-algebraic relations can be put on the $A$'s and the norms can be inherited from $\mathcal{A}$, so that the  elements $(A,\mathcal{V})$ generate the $C^*$-subalgebra $\mathcal{V}$.
This way, the $C^*$-algebra generated by the collective elements $(A,\mathcal{V})$ is the limit object  $\mathcal{A}_{\text{ctx}}$, since for any fixed $\mathcal{V}$ there is a $C^*$-subalgebra $\mathcal{V}$ of $\mathcal{A}_{\text{ctx}}$  and a $*$-homomorphism $\lambda':\mathcal{A}_{\text{ctx}}\rightarrow\mathcal{A}'$  such that the universal relation  $\lambda'\circ\i_\mathcal{V}=\iota_\mathcal{V}$  holds.

As a simple example, consider the Gel'fand spectrum $\Sigma(\mathcal{V})$ of the commutative $\mathcal{V}$. Since the $C^*$-algebra $\mathcal{V}$ is $*$-isomorphic to the algebra $\mathcal{C}(\Sigma)$ of complex-valued functions  on $\Sigma(\mathcal{V})$, we can consider the extension of $\mathcal{V}$ as the extension of  $\mathcal{C}(\Sigma)$. Suppose an extension of $\mathcal{C}(\Sigma)$ is the function algebra on a product $\Sigma(\mathcal{V})\times\Sigma(\mathcal{V}_2)$, then we have
\begin{equation}\label{2222}
\vcenter{\xygraph{
!{<0cm,0cm>;<1cm,0cm>:<0cm,1cm>::}
!{(0,0) }*+{\mathcal{C}(\Sigma(\mathcal{V}))}="a"
!{(1,1) }*+{\mathcal{V}}="b"
!{(3,0) }*+{\mathcal{C}(\Sigma(\mathcal{V})\times\Sigma(\mathcal{V}_2))}="c"
"b":"a"_{\cong}"b":"c""a":"c"
}}\end{equation}
We therefore see that the contextual extension is indeed an extension of the contexts.
 Now the limit object
$\mathcal{A}_{\text{ctx}}$ is  the function algebra $\mathcal{C}(\prod_\mathcal{V}\Sigma(\mathcal{V}))$ on the product as a categorical limit.
If  $\mathcal{V}$ is furthermore modeled on a compact Hausdorff space $X$, then by the Riesz-Markov representation theorem there exists a Borel measure $\mu$ on $X$ such that
\begin{equation}\label{3333}
\varphi(A)=\braket{\Omega|\pi(A)|\Omega}=\int_XA(x)d\mu(x),\quad x\in X
\end{equation}
where $A(x)\in \mathcal{A}$ is a continuous function on $X$. Then $\mathcal{A}_{\text{ctx}}$ is modeled on a product $\prod_\mathcal{V}X_{\mathcal{V}}$ with product topology, and likewise we have
\begin{equation}\label{4444}
\varphi(A)=\int_{\prod_\mathcal{V}X_{\mathcal{V}}}A(x)d\mu(x).
\end{equation}
For two different contexts $\mathcal{V}_1,\mathcal{V}_2$, it is now obvious that $\varphi(A)|_{\mathcal{V}_1}\neq\varphi(A)|_{\mathcal{V}_2}$, as expected.

The situation is slightly different in  algebraic quantum field theories. In the the Haag-Kastler approach \cite{HK64}, one considers the observables  localized in a double cone $U$ in Minkowski spacetime $M$; the algebra of these localized observables form a $C^*$-algebra $\mathcal{A}_U$. For two double cones $U,V$ with $V\subseteq U$, the corresponding algebras satisfy the isotony condition, $\mathcal{A}_V\subseteq\mathcal{A}_U$; if $U$ and $V$ are space-like separated, the Einstein locality becomes the condition $[A,B]=0, A\in\mathcal{A}_U,V\in\mathcal{A}_V$  for algebras. For quantum field theories on the fixed Minkowski spacetime, the  Poincar\'e symmetry on $U$ is translated to the automorphisms $\alpha_g$ on the algebra $\mathcal{A}_U$ with $g$ in the Poincar\'e group. These $\mathcal{A}_U$  form a net of $C^*$-algebras; the total algebra on $M$ can be obtained by the inductive limit
$\mathcal{A}(M)=\overline{\cup_U\mathcal{A}_U}$.

As before, the local contexts for $\mathcal{A}_U$ are the commutative $C^*$-subalgebras $\mathcal{V}_{U,i}$ of $\mathcal{A}_U$. But now it is possible for the measurement contexts to extend to space-like separated double cones. Given a collection of mutually causally-separated double cones $\{U_1,U_2,...,U_k\}$, the measurement contexts could be 
 \begin{equation}\label{4}
\mathcal{V} \equiv\{\mathcal{V}_1,\mathcal{V}_2,...,\mathcal{V}_k\},
 \end{equation}
 each $\mathcal{V}_i$ of which is a commutative subalgebra of $\mathcal{A}_{U_i}$. Such a collection $\{\mathcal{V}_1,\mathcal{V}_2,...,\mathcal{V}_k\}$ of subalgebras generate a tensor product $\mathcal{V}_1\otimes\mathcal{V}_2\otimes...\otimes\mathcal{V}_k$, because the $U_i$'s are given as mutually causally-separated. By considering the injective $*$-homomorphisms $i_\mathcal{V}:\mathcal{V}\rightarrow\mathcal{A}(M)$, we have the contextual extension $\mathcal{A}'(M)$ of $\mathcal{A}(M)$. The $\mathcal{A}'(M)$ needs to be consistent with the extra structures in $\mathcal{A}(M)$: Firstly, the extension should have the automorphisms implementing the Poincar\'e symmetry. Let $\alpha_g$ be the automorphism of $\mathcal{A}(M)$, then there should exist the automorphism $\alpha'_g$ of $\mathcal{A}'(M)$ such that $\alpha'_g\iota_\mathcal{V}=\iota_{g(\mathcal{V})}\alpha_gi_\mathcal{V}$ with $\iota_\mathcal{V}:\mathcal{V}\rightarrow\mathcal{A}'(M)$. Secondly, the contextual extensions can be localized to the double cones as $\mathcal{A}'_U$ and fulfill the defining conditions of the net of local $C^*$-algebras. In this sense, the  $\mathcal{A}'(M)$ is a {\it local} contextual extension.

In the category $\mathsf{A}_L$ of local contextual extensions with injective $*$-homomorphisms as morphisms, we can consider the cone as in \eqref{1} and the the limit object $\mathcal{A}_{\text{lctx}}(M)$. Again, if we consider the Gel'fand spectra $\Sigma(\mathcal{V})$ of $\mathcal{V}$ and extend the function algebra $\mathcal{C}(\Sigma(\mathcal{V}))$, the limit  object $\mathcal{A}_{\text{lctx}}(M)$ is the function algebra on the product of the spectra, $\mathcal{C}(\prod_\mathcal{V}\Sigma(\mathcal{V}))$.
 
 The above construction can be similarly applied to the more general case of locally covariant quantum field theories  \cite{BFV03}. A locally covariant quantum field theory is a covariant functor 
  \begin{equation}\label{F}F:\mathsf{M}\rightarrow\mathsf{A},\end{equation}
 where $\mathsf{M}$ is  the category of globally hyperbolic, oriented and time-oriented spacetime manifolds with isometric embeddings as morphisms, and $\mathsf{A}$ is the category  of unital $C^*$-algebras for algebraic quantum theory. Given a spacetime manifold $(M,g)\in\mathsf{M}$, the generally noncommutative $C^*$-algebra $F(M,g)$ can have commutative subalgebras $\mathcal{V}$'s corresponding to the bounded open subsets $(U,g_U)$ of $(M,g)$: we have the commutative diagram
 \begin{equation}\label{5}
\vcenter{\xygraph{
!{<0cm,0cm>;<1cm,0cm>:<0cm,1cm>::}
!{(0,0) }*+{F(U,g_U)}="a"
!{(0,1.3) }*+{(U,g_U)}="b"
!{(2,0) }*+{F(M,g)}="c"
!{(2,1.3) }*+{(M,g)}="d"
"b":"a"_{F}"b":"d"^{\iota_{M,U}}"a":"c"^{i}"d":"c"^{F}
}}\end{equation}
where $\iota_{M,U}:(U,g_U)\rightarrow(M,g)$ is the isometric embedding induced from $\mathsf{M}$.
 We have   therefore as before the injective $*$-homomorphisms $i_\mathcal{V}:\mathcal{V}\rightarrow F(M,g)$ as inclusions with $\mathcal{V}$ being in the composite sense of \eqref{4}. By taking these $i_\mathcal{V}$ as the legs of a cone over a diagram $F$, we can define the contextual extension of the algebra $F(M,g)\in\mathsf{A}$ as in \eqref{1}.
 \subsection{Contextual extensions in topos quantum theory }
 As we have seen, the contextual extensions reviewed above can be formulated in a general sense as cones in a category.
 Let us consider here some examples of contextual extensions in this general sense in topos quantum theory.
 
 Consider the category $\mathsf{V}$ of commutative $C^*$-subalgebras of a $C^*$-algebra $\mathcal{A}$ as above. If $\mathcal{A}$ is a von Neumann algebra, the category
 $\mathsf{V}$ corresponds to the {\it context category} in the contravariant topos quantum theory. Let $\underline{\Sigma}_V$ be the
 Gel'fand spectrum of $V$. 
The {\it spectral presheaf} $\underline{\Sigma}$ on the {context category} $\mathsf{V}$ is the contravariant functor acting on objects and morphisms of  $\mathsf{V}$ respectively as
\begin{equation}
\underline{\Sigma}:V\rightarrow\underline{\Sigma}_V; \Bigl(i_{V_2V_1}:V_1\rightarrow V_2\Bigr)\rightarrow \Bigl(\underline{\Sigma}(i_{V_2V_1}):\underline{\Sigma}_{V_2}\rightarrow\underline{\Sigma}_{V_1}\Bigr)
\end{equation}
where $V_1\subset{V_2}$ for $V,V_1,V_2\in\text{Ob}(\mathsf{V}(H))$. The category of spectral presheaves form a topos $\tau_\mathsf{V}$. Now we trivially have the cone in $\tau_\mathsf{V}$ over the diagram defined by the functor $\underline{\Sigma}:\mathsf{V}\rightarrow\tau_\mathsf{V}$,
\begin{equation}\label{7}
\vcenter{\xygraph{
!{<0cm,0cm>;<1cm,0cm>:<0cm,1cm>::}
!{(0,0) }*+{\underline{\Sigma}_{V_1}}="a"
!{(1,1) }*+{\underline{\Sigma}_{V_1}}="b"
!{(2,0) }*+{\underline{\Sigma}_{V_2}}="c"
"b":"a"_{\text{id}}"b":"c""c":"a"^{\underline{\Sigma}(i)}
}}\end{equation}
We see that the Gel'fand spectrum $\underline{\Sigma}_{V_2}$ is a contextual extension of $\underline{\Sigma}_{V_1}$, while the $\underline{\Sigma}_{V_1}$ is a coarse-graining of $\underline{\Sigma}_{V_2}$.
By definition, the topos $\tau_\mathsf{V}$ is a cartesian closed category, so the limit object of the cone is a finite product of Gel'fand  spectra. A proposition in the quantum theory is therefore always a subobject of the limit object.

To obtain a nontrivial  cone over the category of noncommutative algebras, we change to the covariant topos quantum theory. In the covariant theory, the Gel'fand spectra $\underline{\Sigma}_\mathcal{V}$ of the commutative subalgebras $\mathcal{V}\equiv\underline{\mathcal{A}}$ of a $C^*$-algebra $\mathcal{A}$ are organized into a locale in a topos $\tau$. Then the Gel'fand transform associates a locale map to the commutative $\mathcal{V}$ internal to $\tau$ by
\begin{equation}
\hat{\mathcal{V}}:\underline{\Sigma}_\mathcal{V}\rightarrow\underline{\mathsf{R}}
\end{equation}
where $\underline{\mathsf{R}}$ is the Dedekind real-number object in $\tau$. This locale map can be generalized to the (covariant) daseinisation
\begin{equation}
\underline{\delta}(\mathcal{V}):\underline{\Sigma}_\mathcal{V}\rightarrow\underline{\mathsf{IR}}
\end{equation}
where $\underline{\mathsf{IR}}$ is the interval domain in $\tau$. This $\underline{\delta}(\mathcal{V})$  deviates from  the Gel'fand spectra of commutative $\mathcal{V}$, and hence it describes the general elements in $\mathcal{A}$. 
By switching to the frames opposite to locales, we have the following cone over the diagram indexed still by $\mathsf{V}$,
\begin{equation}\label{10}
\vcenter{\xygraph{
!{<0cm,0cm>;<1cm,0cm>:<0cm,1cm>::}
!{(0,0) }*+{O(\underline{\mathsf{IR}})}="a"
!{(1,1) }*+{O(\underline{\Sigma}_\mathcal{V})}="b"
!{(2,0) }*+{O'(\underline{\mathsf{IR}})}="c"
"b":"a"_{\underline{\delta}(\mathcal{V})}"b":"c"^{\underline{\delta}'(\mathcal{V})}"a":"c"_i
}}\end{equation}
where the $O$s are frames and the $i$ is a frame homomorphism. By definition a frame is complete lattice, so there exists a limit object as the $\sup$ of the frame if $O'(\underline{\mathsf{IR}})$ is the extension by joins.
 
 We thus see that the contextual extensions naturally exist in topos quantum theory, both in the states as spectral presheaves and in the observables as daseinisations. This is not surprising, because the topos quantum theory is formulated to describe the contextual structures (or classical faces) of quantum theory by construction. 

\section{Contextual extensions of   quantum gravity  models }\label{III}
In this section we turn to the contextual extensions of quantum gravity models. We first consider the nonperturbative models of quantum gravity, in particular the algebraic formulation of group field theory \cite{KOT18}. We then show a streamlined formal analogy of contextual extension in the BV formulation of  perturbative  quantum gravity  \cite{BFR16}. Finally, we comment on the topos structure of  contextually extended quantum gravity models.

\subsection{Nonperturbative models}
Let us consider the algebraic formulation of group field theory \cite{KOT18}. To this end, it is helpful to introduce the quantum geometry of group field theory according to nonperturbative loop quantum gravity: A (gauge-invariant) group field $\phi(\{g_i\})$ corresponds to a  quantum polyhedron, so the $g_i$s are group elements in the gauge group of loop quantum gravity. By defining the  vacuum  $\ket{0}$ as the state without any quantum geometric excitation, we have that a  group field $\phi$ and its conjugate $\phi^\dag$ annihilates and  respectively  creates a quantum polyhedron, i.e.
$
\phi(\{g_i\})\ket{0}=0$ and $\phi^\dag(\{g_i\})\ket{0}=\ket{\{g_i\}}$.
They satisfy the {bosonic} canonical commutation relations (CCR)
$
[\phi(\{g_i\}),\phi(\{g_j\})^\dag]=\delta(\{g_i\},\{g_j\}),
$
and also generate
  a bosonic Fock space
\begin{equation}\label{222}
\mathcal{H}_{\text{Fock}}=\bigoplus_{N>0}\text{sym}\mathcal{H}^{\otimes N}
\end{equation}
where $\mathcal{H}_{(n)}=\otimes_{i=1}^n\mathcal{H}_i$ is the Hilbert space of a single polyhedron with $n$ faces (and $i=1,..,n$ in $g_i$). As in  second quantization, the group fields can be further smeared by test functions $f$,
\begin{equation}\label{23}
\Psi(f)=\int_{G^{\times n}}\prod_{i=1}^nd\mu(g_i){f(\{g_i\})} \phi(\{g_i\})
\end{equation}
where  $\mu$ is the Haar measure on the gauge group $G$. 
 These field operators $\Psi$ inherits the CCR of $\phi$ with the $\delta$-function replaced by the  $L^2$ inner product of the test functions
\begin{equation}\label{24}
[\Psi(f),\Psi(f')^\dag]=\int_{G^{\times n}}\prod_{i=1}^nd\mu(g_i){f(\{g_i\}) }\overline{f'(\{g'_i\})}\equiv(f,f').
\end{equation}
The inner product $(f,f')$ appears in the Weyl  commutation relation 
\begin{equation}W_{(f)}\cdot W_{(f')}=e^{-\frac{i}{2}\text{Im}(f,f')}W_{(f+f')}
\end{equation}
for the Weyl elements  $W_{(f)}=\exp\{\frac{i}{\sqrt{2}}\Psi(f)+\Psi^\dag(f)\}$. This Weyl algebra $\mathcal{W}$  is  also a
$C^*$-algebra \cite{KOT18}. The $N$-body field operators and the corresponding Weyl algebras can be constructed as in second quantization.

To consider the contextual extensions we change to the category-theoretic framework. We have, on the one hand, the category $\mathsf{PH}$ of the direct sums of single-polyhedron Hilbert spaces $\mathcal{H}_{(n)}$ with a fixed $n$ where the morphisms are the 
  subspace inclusions, and on the other hand, the category $\mathsf{W}$ of Weyl $C^*$-algebras with $*$-homomorphisms as morphisms. The second quantization of group fields recalled above is then a functor 
\begin{equation}
S:\mathsf{PH}\rightarrow\mathsf{W}.
\end{equation}   Let $\mathcal{V}_G$ be commutative $C^*$-subalgebra of $\mathcal{W}$; we define these commutative subalgebras as the ``contexts''. In a ``context'', the commuting Weyl elements correspond to the group field operators with $\text{Im}(f,f')=0$, which means $\mathcal{V}_G$ is no longer a nontrivial  Weyl algebra. Let $i_\mathcal{V}:\mathcal{V}_G\rightarrow\mathcal{W}$ be the $*$-homomorphism as subalgebra inclusion.
We can formulate the cones  in $\mathsf{W}$ indexed by $\mathsf{H}$ over the functor $S$,
\begin{equation}\label{15}
\vcenter{\xygraph{
!{<0cm,0cm>;<1cm,0cm>:<0cm,1cm>::}
!{(0,0) }*+{S(\oplus^k\mathcal{H})}="a"
!{(1,1) }*+{\mathcal{V}_G}="b"
!{(2,0) }*+{S(\oplus^l\mathcal{H})}="c"
"b":"a"_{i_\mathcal{V}}"b":"c""a":"c"_{s_{kl}}
}}\end{equation}
where $s_{kl}$ with $k<l$ is the the $*$-homomorphism induced from the inclusion in $\mathsf{H}$. Notice that now the cone \eqref{15} is indexed by $\mathsf{PH}$, but by smearing the group fields with the test functions satisfying $\text{Im}(f,f')=0$ we can still consider the ordering in  $\mathsf{PH}$ as coming from a context category $\mathsf{V}$. 

 The  extension $S(\oplus^l\mathcal{H})$ as a Weyl $C^*$-algebra of group field theory contains more quantum-geometric excitations than $S(\oplus^k\mathcal{H})$; conversely, $S(\oplus^k\mathcal{H})$ is a coarse-graining of $S(\oplus^l\mathcal{H})$ by eliminating quantum polyhedra. The limit object of the limit of \eqref{15} contains all possible contextual extensions and hence it is the finest algebra corresponding to the Fock space \eqref{222}.

If   we alternatively  consider the category $\mathsf{FH}$ of Fock spaces as in \eqref{222} but with different $k$ whose morphisms are the subspace inclusions of each single-polyhedron factors, e.g. $i_\mathcal{H}:\mathcal{H}_{(k)}\rightarrow\mathcal{H}_{(l)}$ with $k<l$. In this case, we can consider the functor $W:\mathsf{FH}\rightarrow\mathsf{W}$, then we have as in  \eqref{15} the morphism $w_{kl}:W(\oplus\mathcal{H}_{(k)})\rightarrow W(\oplus\mathcal{H}_{(l)})$ in $\mathsf{W}$. We see that $W(\oplus\mathcal{H}_{(k)})$ is a coarse-graining of $W(\oplus\mathcal{H}_{(l)})$ by forgetting the ``faces'' uniformly. But now the limit object contains Hilbert spaces corresponding to polyhedra with a large number of faces. With a change of perspective, we can take the large number of faces with small spins as the large values of spins on a small number of faces; in this sense, the limit object describes the situation  similar to the semiclassical limit of spin foam models.

Group field theory  is an omnibus of structures from various modern models of quantum gravity.
 Next, we  consider similarly  the spin foam models as another example.

 Instead of algebraic category, let us consider the lattice category of spin foams \cite{Guo18}: Let $\mathcal{K}$ be the kinematic Hilbert space on the spatial 3-boundary as given by loop quantum gravity.  Recall that in the language of lattice theory, the measurement contexts are the Boolean  sub-$\sigma$-algebras $\mathcal{P}(\mathcal{K})$ of the Hilbert lattice $\mathcal{L}(\mathcal{K})$ \cite{Gud70}. The $\mathcal{L}(\mathcal{K})$
 can be enriched  by a complete meet lattice of probability functions (as induced by the Boolean  sub-$\sigma$-algebras $\mathcal{P}(\mathcal{K})$) to form a quantaloid $\mathcal{Q}(\mathcal{K})$. Using $\mathcal{Q}(\mathcal{K})$ we can induce the tensor products of two kinematic Hilbert spaces, which is otherwise undefinable using lattices.  Let $c:\mathcal{P}(\mathcal{K})\rightarrow\mathcal{Q}(\mathcal{K})$ be the inclusion map that preserves the lattice order. 
 Then the contextual extensions can be considered on the $\mathcal{Q}$s:
 \begin{equation}\label{19}
\vcenter{\xygraph{
!{<0cm,0cm>;<1cm,0cm>:<0cm,1cm>::}
!{(0,0) }*+{\mathcal{Q}(\mathcal{K})}="a"
!{(1,1) }*+{\mathcal{P}(\mathcal{K})}="b"
!{(2,0) }*+{\mathcal{Q}(\mathcal{K}')}="c"
"b":"a"_{c}"b":"c"^{c'}"a":"c"_{C}
}}\end{equation}
If $\mathcal{Q}(\mathcal{K}')$ is the evolution of $\mathcal{Q}(\mathcal{K})$ according to the transition rules of a spin foam model, then $C$ should be the  map corresponding to the  tensor product $\mathcal{K}^*\otimes\mathcal{K}$ defined with the help of the quantaloids. 

 We can therefore consider the category $\mathsf{Q}$ of quantaloids in analogy to the bordism category: the objects are the quantaloids $\mathcal{Q}$ and the morphisms are those $C$s. On the other hand, we have the category $\mathsf{S}$ of simplicial foams, i.e. the category with as objects the spin networks living on the 3-boundaries and as morphisms the cobordisms between the branched coverings of the 3-boundaries. Then a spin foam model is a functor 
 \begin{equation}
 SF:\mathsf{S}\rightarrow\mathsf{Q}.
 \end{equation}
 In this sense, the  $\mathcal{Q}(\mathcal{K})$  can be understood again as the coarse-graining of $\mathcal{Q}(\mathcal{K}')$ by changing the simplicial complexes defining the foams.

\subsection{Perturbative models}
In the previous subsection we did not the symmetry conditions that the contextual extensions should satisfy. This is because the symmetries usually emerge, instead of being imposed, in these constructive models. However, in perturbative quantum gravity the symmetries of the background spacetime have to be taken into consideration.

 Let us consider the background-independent BV effective theory approach to perturbative quantum gravity  \cite{BFR16}. In the BV approach, the infinitesimal diffeomorphism of spacetime is reexpressed through the ghost fields $c$; we have the collective field data $\varphi=(g,c,\bar{c}_\mu,b_\mu),\mu=0,1,2,3$, where $g$ is the metric fields of classical spacetime $M$, $\bar{c}_\mu$ and $b_\mu$ are the antighosts and scalars given by the gauge-fixing. These $\varphi$s together with the BV operator $s$ form the BV algebra $\mathcal{BV}$; further, the BV algebras $\mathcal{BV}$ and 
 the ((-1)-shifted) Poisson brackets  on $\mathcal{BV}$ together form a BV manifold $BV(M)$. 
 In perturbative quantum gravity, one has a background metric $g_0$ and considers the perturbed field $\varphi_\lambda=(g_0+\lambda h,\lambda c,\lambda\bar{c}_\mu,\lambda b_\mu)$. The $g_0$  (i.e. $\lambda=0$) part corresponds to the ``classical'' part $S_0$ of the total action $S$, and the perturbation part is the rest of $S$ which is expanded in orders of the parameter $\lambda$. 
 
 The BV quantization endows the algebra $\mathcal{BV}$ with a noncommutative star product (with a deformation parameter $\hbar$) as in deformation quantization. Let us denote this  algebra with noncommutative star product by $\mathcal{A}_\star$.
 In the classical limit $\hbar\rightarrow0$ the noncommutatively deformed product reduces to the commutative product, and the effective action is reduced to  $I_0$   part of  $I=I_0+I_1\hbar+I_2\hbar^2+\ldots$.  $I_0$ satisfy the classical master equation $Q_0I_0+\frac{1}{2}\{I_0,I_0\}=0$, which means the classical infinitesimal gauge symmetry in the BRST sense but not the full quantum gauge symmetry.
  The ``classical''  action is then  $S_0=\omega+I_0$ where $\omega$ is the free action given by the symplectic pairing.
  
  In the BV quantization of perturbative quantum gravity,  the total effective action $I$ containing the perturbation part needs to be renormalized. The general renormalization in BV formalism \cite{Cos11} considers the the renormalized effective actions $I[r]$ (as the solutions of the renormalized quantum master equation $(Q+\hbar s_{K_r})e^{I[r]/\hbar}=0$ with the smooth kernel $K_r$ homologous to the singular Poisson kernel $K_0$). Let $\mathcal{BV}[r]$ be the BV algebra corresponding to the renormalized effective action $I[r]$.
  The commutative  subalgebra  $\mathcal{BV}_0$ of $\mathcal{A}_\star$ corresponding to $I_0$ is  the  double limit $\hbar\rightarrow0,r\rightarrow0$ of  $\mathcal{BV}[r]$. However, the  $\mathcal{BV}_0$ is unique if the background is fixed. We then consider the renormalized  BV algebras $\mathcal{BV}[r]$ before deformation quantization ($\hbar\rightarrow0$) as the ``contexts''.
 
Let $\mathsf{BV}$ be the category of $\mathcal{BV}[r]$s with the homotopic renormalization group flow as the morphisms; let $\mathsf{A}_\star$ be the category of the quantized noncommutative algebras $\mathcal{A}_\star$s with algebraic homomorphisms as morphisms. The BV quantization is then a functor $B:\mathsf{BV}\rightarrow\mathsf{A}_\star$,.
We see that a perturbative quantum gravity model is a  locally covariant quantum field theory with the functor $F$ factorized through the category of BV algebras (or BV manifolds),
\begin{equation}
F:\mathsf{M}\rightarrow\mathsf{BV}\xrightarrow{B}\mathsf{A}_\star.
\end{equation}
Suppose we have in general the injective homomorphism $e:\mathcal{BV}[r]\rightarrow B(\mathcal{BV}[r_1])$ from a renormalized BV algebra into the commutative subalgebra of $B(\mathcal{BV}[r_1])$.
We can consider the cones in $\mathsf{A}_\star$ over $B$,
\begin{equation}\label{22}
\vcenter{\xygraph{
!{<0cm,0cm>;<1cm,0cm>:<0cm,1cm>::}
!{(0,0) }*+{B(\mathcal{BV}[r_1])}="a"
!{(1,1) }*+{\mathcal{BV}[r]}="b"
!{(3,0) }*+{B(\mathcal{BV}[r_2])}="c"
"b":"a"_{e}"b":"c"^{\epsilon}"a":"c"_{R}
}}\end{equation}
where
 $R$ is induced from the renormalization group flow between the $\mathcal{BV}[r]$s. 
Therefore the extension $B(\mathcal{BV}[r_2])$ can be understood as a renormalization transform of $B(\mathcal{BV}[r_1])$. The limit of the cone \eqref{22} is then a renormalization group fixed point to which all $B(\mathcal{BV}[r])$ can flow.

\subsection{Topos of quantum  gravity models}
The paradigm we have been considering is the following. A quantum gravity model is a functor $F:\mathsf{X}\rightarrow\mathsf{T}$ from the category $\mathsf{X}$ of configurations to the category $\mathsf{T}$ of theories. $\mathsf{X}$  is equivalent to or  related in a certain way to a context category $\mathsf{V}$ and share a preorder structure.
Then the cone with commutative summit $T_c\in\mathsf{T}$ over the functor $F$ 
\begin{equation}\label{23}
\vcenter{\xygraph{
!{<0cm,0cm>;<1cm,0cm>:<0cm,1cm>::}
!{(0,0) }*+{T}="a"
!{(1,1) }*+{T_c}="b"
!{(2,0) }*+{T'}="c"
"b":"a""b":"c""a":"c"
}}\end{equation}
defines the patterns of contextual extensions $T'$ of a theory $T\in\mathsf{T}$. Such extensions are called {\it contextual} because the cones \eqref{23} are (effectively) indexed by a context category $\mathsf{V}$.
The ordering  in $\mathsf{T}$ inherited from $\mathsf{X}$ (or $\mathsf{V}$)  can be interpreted in various ways, e.g. the coarse-graining.

So far, we have taken the existence of contexts for granted, but it is quite possible for the models/theories to be noncontextual. Given two noncontextual theories $T_1,T_2\notin\text{Obj}(\mathsf{T})$, we  consider their contextual extensions $\tilde{T}_1,\tilde{T}_2$ in the ``nothing-to-something'' sense  by adding different numbers of contexts to $T_1,T_2$. Then $\tilde{T}_1,\tilde{T}_2$ and a common context $T_{c}$ of them can be organized into a contextual extension as in \eqref{23} such that $\tilde{T}_1,\tilde{T}_2\in\text{Obj}(\mathsf{T})$; in this sense, the contexts equip the theory space with a particular ordering which otherwise does not exist. 
Mathematically, if the category $\mathsf{X}$ is   a locale, the  preorder  structure of $\mathsf{X}$ is transferred to $\mathsf{T}$. Then   $\mathsf{T}$   is a site $(\mathsf{T},J)$ with  the Grothendieck topology $J$ given by the locale $\mathsf{X}$. This way, we have the Grothendieck topos $\mathsf{Sh}(\mathsf{T},J)$. More generally, we have the elementary topos $[\mathsf{T}^{\text{op}};\mathsf{Set}]$ as the category of presheaves if we do not impose the Grothendieck topology $J$.

  The topos $[\mathsf{T}^{\text{op}};\mathsf{Set}]$, however, does not exactly conform to the topos quantum theory. The major difference is that, this topos structure lives at the level of {\it total theories},  instead of at the  level of ``states''  in topos quantum theory. In physics terms, the topos structure is now in the theory space, while in topos quantum mechanics the topos is in  the state space.
  Since a theory $T$  itself can be described by a category, we see that the contextual extensions give rise to a higher categorical structure. This higher extension seems to be a necessarily step towards the many-body or field-theoretic generalization of topos quantum mechanics. (Cf. \cite{Vei17} for the philosophical arguments on this point.) 
  
In the special case where a theory $T_0$ is extended by the products of theories as the product  category $\mathsf{T}=\prod_aT_a$, the theory topos $[\mathsf{T}^{\text{op}};\mathsf{Set}]$ with the contextual ordering is  known as the {\it temporal topos} in \cite{Kat04}.   Unlike the compatible observables in a given context, this ``temporal'' interpretation indicates that the contextual extensions of $T_0$ generically  contain mutually incompatible observables. To test classical realisms in this case, the corresponding realism inequalities then should be of the ``temporal''  Legget-Garg type  at the level of theories . However, it is still possible to formulate the hybrid realism inequalities for both the compatible observables in given contexts and the incompatible observables from different theories $T$s \cite{DASH13}.\footnote{In the case of the  algebras over the  product of Gel'fand spectra, i.e. $\mathcal{C}(\prod_a\Sigma_a)$ as in \eqref{2222}, we can write the correlation between two sets of observables $\{A_i\},\{B_j\}$ via the integral representation \eqref{4444} as
\begin{equation}\label{24}
\braket{A_iB_j}=\int_{\prod_\mathcal{V}X_{\mathcal{V}}}A_i(x)B_j(x)d\mu(x).
\end{equation}
Notice that the $\{A_i\},\{B_j\}$ here are not required to be in the same context. The we obtain the hybrid realism inequalities by replacing the products with the correlations  \eqref{24} in  the Roy-Singh inequality,
\[
\sum_{p=1}^q\Bigl(\sum_{i=1}^{m_p}(\pm A_i^{(p)})+\sum_{j=1}^{n_p}(\pm B_j^{(p)})\Bigr)^2\geqslant q
\]
where $m_p+n_p$ is an odd number.}


\section{Conclusion}\label{IV}
We have presented a simple approach to the contextual extensions of quantum gravity models. This approach generalizes the contextual extensions of $C^*$-algebras to the categorical framework, and can be applied in both the nonperturbative and the perturbative modern models of quantum gravity. 

This paper only considers the general categorical structures related to contextual extensions. To see whether these general structures are useful for explicit model-buildings, we need to consider more detailed quantum gravity models.
We also admit that the contextual extension is not  the best way to capture the physics of contextuality, since the contexts are only formally defined as the commutative subalgebras without specifying the measurements. 
If one can find a well-defined operational  quantum-measurement scheme in quantum gravity, the noncontexuality inequality remains a good way to distinguish contextuality from noncontexuality.

We give a final remark that the contextual extensions considered here will be the same  as the generalized contexts formalism in quantum theory  \cite{LV09} if the contextual extensions happen at different times. 
\begin{acknowledgments}
This work is partially supported by the National Natural Science Foundation of China through the
Grant Nos. 11875006 and 11961131013.
\end{acknowledgments}


\end{document}